\documentclass{article}
\usepackage{graphicx}
\usepackage[english]{babel}
\usepackage{srcltx}
\usepackage{amsmath}
\usepackage{amssymb}
\usepackage{amsfonts}
\usepackage{latexsym}
\usepackage{textcomp}
\usepackage{appendix}
\usepackage{multirow}
\usepackage{booktabs}
\usepackage{rotating}
\usepackage{url}
\usepackage{array}
\usepackage{marvosym}
\usepackage{afterpage}
\usepackage{pdflscape}
\usepackage[table,svgnames]{xcolor}
\usepackage{graphics}
\usepackage{graphicx}
\usepackage{caption}
\usepackage{multirow}
\usepackage{booktabs}
\usepackage{standalone}
\usepackage{tabulary}
\usepackage{adjustbox}
\usepackage{booktabs}
\usepackage{pdflscape,afterpage,caption}
\usepackage{longtable}
\usepackage[square,numbers]{natbib}
\usepackage{subfig} 
\usepackage{lineno,hyperref}
\usepackage{authblk}
\modulolinenumbers[5]
\usepackage[margin=1.25in]{geometry}

\title{Covid-19 impact on cryptocurrencies: evidence from a wavelet-based Hurst exponent}

\author[1]{M. Bel\'en Arouxet}
\author[2]{Aurelio F. Bariviera\thanks{Corresponding author.\texttt{aurelio.fernandez@urv.cat} }}
\author[3,4]{Ver\'onica E. Pastor}
\author[4]{Victoria Vampa}
\affil[1]{\footnotesize Universidad Nacional de La Plata, Facultad de Ciencias Exactas, Centro de Matemática de La Plata, Argentina}
\affil[2]{\footnotesize Universitat Rovira i Virgili, Department of Business, Av. Universitat 1, 43204 Reus, Spain}
\affil[3]{\footnotesize Universidad de Buenos Aires, Facultad de Ingenier\'ia, Departamento de Matem\'aticas}
\affil[4]{\footnotesize Universidad Nacional de La Plata, Facultad de Ingenier\'ia, Departamento de Ciencias B\'asicas, Argentina}

\begin{document}
\maketitle

\begin{abstract}
Cryptocurrency history begins in 2008 as a means of payment proposal. However, cryptocurrencies evolved into complex, high yield speculative assets. Contrary to traditional financial instruments, they are not (mostly) traded in organized, law-abiding venues, but on online platforms, where anonymity reigns. This paper examines the long term memory in return and volatility, using high frequency time series of eleven important coins. Our study covers the pre-Covid-19 and the subsequent pandemia period. We use a recently developed method, based on the wavelet transform, which provides more robust estimators of the Hurst exponent. We detect that, during the peak of Covid-19 pandemic (around March 2020), the long memory of returns was only mildly affected. However, volatility suffered a temporary impact in its long range correlation structure. Our results could be of interest for both academics and practitioners.

{\bf Keywords:} cryptocurrencies; Hurst exponent; wavelet transform; Covid-19
\end{abstract}

\section{Introduction}
Cryptocurrencies have become one of the most traded financial assets in the last decade. In order to put their importance into perspective, two of the most important stock exchanges in the world, the New York Stock Exchange and Nasdaq, report an average of \$30 billions and \$85 billions in daily volume, respectively. Over the last six months, daily transactions of cryptocurrencies varied between \$5 bilions and \$31 billions, depending on the day. As a consequence, cryptocurrencies have been receiving increasing interest from both academics and practitioners. As a new object of study, it poses several challenges. One of them is to examine the statistical properties of the price generating process. According to the Efficient Market Hypothesis (EMH), the price of any speculative asset must convey all available information \citep{Fama70}. In particular, the weak version of the EMH states that the current price includes all the information contained in the series of past prices. As a consequence of the non-arbitrage possibility, price returns time series should follow a random process with no memory. In particular, it is excluded the possibility of long-term memory, as it could allow for profitable trading strategies. 

The EMH, specially regarding the presence of long-term memory has been subject to debate since the work by Mandelbrot \cite{Mandelbrot1963}. It has been extensively studied in developed and emerging stock markets \cite{Barkoulas96,Panas01,CTcauses}, in fixed income markets \cite{Carbone04,BaGuMa12,BaGuMa14}, interest rates \cite{GARCIN2017462,SENSOY201385,CajueiroInterestRates}, and exchange rates \cite{Matsushita,SouzaCT,YANG2019734}, among other financial time series. 

Studies related to cryptocurrencies are much more recent. Specially on these days it is relevant to investigate how markets react to such a global event as Covid-19 pandemic. 

Early papers studying the informational efficiency of Bitcoin time series \citep{Urquhart2016,Nadarajah2017} find that the returns (and some power transformations) had been informational inefficiency, but the also report a trend toward a more efficient behavior. Shortly after these papers, \cite{Bariviera2017} confirms the diminishing memory in daily returns, along with a highly persistent component in daily volatility. Such persistence justifies the use of GARCH-type models in volatility, as proposed by \cite{KATSIAMPA20173,Katsiampa2018}. 

One-time-only events such as a hughe price crash or a pandemic, could alter the stochastic process governing returns and volatilities. In this line, \cite{Bourietal2017} finds that before the big price crash of 2013, Bitcoin volatility was asymmetric\footnote{Assymetry in volatility means that the reaction of volatility to unexpected positive and negative changes is not the same.} in the opposite way of the traditional assets, whereas this asymmetry is not found after the price crash. Similarly, a global event such as Covid-19 pandemic, could have non-trivial effects on volatility and returns. Consequently, it begins to be an area of interest for researchers. For example, \cite{CorbetHou2020} reports significant growth in both returns and volumes traded in large cryptocurrencies, and \cite{Goodell2020} affirms that levels of Covid-19 caused a rise in Bitcoin prices.

\begin{figure}[!htbp]
\centering
\includegraphics[width = 0.9\textwidth]{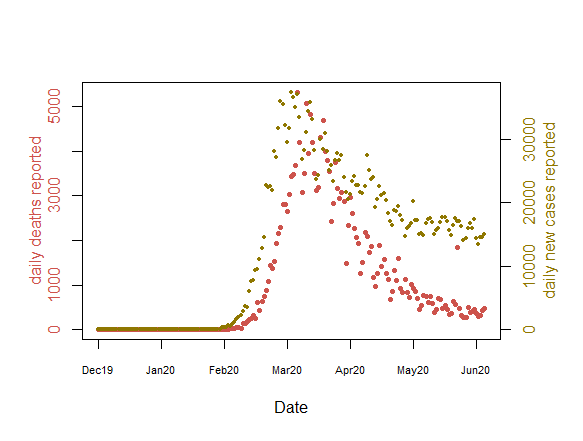}
\caption{Number of new cases and deaths reported per day for European countries. Source: European Centre for Disease Prevention and Control \cite{ecdc}.}
\label{fig:covidcases}
\end{figure}

According to the World Health Organization (WHO) \cite{who}, Covid-19 pandemic originates in the People's Republic of China. On December 31st, WHO's country office gathers information issued by the Wuhan Municipal Health Commission reporting cases of `viral pneumonia'. At the beginning, the virus seemed to be circumscribed to that region. On January 24th. France informed of some cases from people who had been in the Wuhan region, constituting the first confirmed cases in Europe. However, infections and deaths associated to Covid-19 remained at relatively very low levels during January and February. As displayed in Figure \ref{fig:covidcases}, daily number of infected and dead people in Europe rocketed in March, and showed a diminishing trend by the end of that month. Europe has been the first region (after the inception in China) to suffer the pandemic, with its most deadly effect concentrated in a short time frame. Consequently, we would like to investigate if such a sudden public health problem has had an impact on the cryptocurrency market.

In this line, the aim of this paper is to study the long memory profiles of returns and volatilities of eleven cryptocurrencies, during a period spanning before and after the inception of the pandemic event. We contribute to the literature in multiple ways: (i) we propose a new method that has not been applied before to compute the Hurst exponent in cryptocurrencies time series; (ii) we study a set of the eleven most important cryptoassets at high frequency; (iii) we discuss the effect of Covid-19 pandemic on returns and volatilities. 

This remaining of the article is organized as follows: Section \ref{sec:methods} presents the methodology used in the paper; Section \ref{sec:application} describes the data used and discusses the empirical findings; and Section \ref{sec:conclusions} concludes.

\section{Methods \label{sec:methods}}

Harold Edwin Hurst was a British engineer, whose name is intrinsically connected to the study of long range dependence in time series. His original method was presented in a series of papers in the 1950s \citep{Hurst1951,Hurst1956a,Hurst1956b,Hurst1957}. Although it was originally formulated for the resolution of an specific hydrological problem, it turned out to have more universal applications
in the field of time series analysis. The Hurst exponent describes the persistent or anti-persistent character of a time series, arising from its long-range memory. 

The presence of long-range memory is compatible with the fractional Brownian motion model postulated in \cite{MandelbrotVanNess1968,Mandelbrot1968b}. It was precisely Benoit Mandelbrot \cite{Mandel72} who proposed in the early 1970s the use of the fractional Brownian motion model in economics \cite{Mandelbrot1963}. In a series of papers Mandelbrot and coworkers propose a generalization of the standard Brownian motion, which allows for the presence of long memory \citep{MandelbrotVanNess1968,Mandelbrot1968b,Mandelbrot1969}. 

Undoubtedly, finding an accurate measure of long range dependence is a desirable goal in time series analysis. The original $R/S$ method developed by Hurst, has some drawbacks, as it is biased towards finding spurious long memory. This situation triggered the development of alternative methods (e.g. aggregated variance approach \cite{Beran1989}, Higuchi method \cite{HIGUCHI1988277}, Detrended Fluctuation Analysis \cite{Peng1994}, etc.) to find better estimators for long-range dependence.  The review conducted in \cite{serinaldi} provides a comprehensive guide for the proper use of the different methods according to the signal characteristics. 

Previous works acknowledges that, for an arbitrary artificial time series, the wavelet method does not only find a more accurate $H$ value than the R/S method, but also that it is not necessary to assess \textsl{a priori} the series stationarity \citep{serinaldi}. Consequently, in this work we use a wavelet-based method to estimate the Hurst exponent. 

The continuous wavelet transform allows to decompose the time series in the time-frequency domain, and is defined as:
\begin{equation}\label{eq-tw}
W_{\psi}f(a,b) =\frac{1}{\sqrt{a}}\int^{\infty}_{-\infty} f(s)\psi\left(\frac{s-b}{a}\right) ds \,\,\,\,\,\,\,\,\text{for} \,\,\,\, a>0
\end{equation}
where $a$ is the scale parameter, $b$ is the shift parameter and $\psi$ is the mother wavelet. Hereinafter, we will refer to this equation as $W_{a,b}$.

If the time series has $H-$self-affinity, i.e. if the time series satisfies a power law of the kind $X(ct)\approx c^H X(t)$, the variance of the wavelet transform (Eq. (\ref{eq-tw})) will be asymptotically affected by a scale parameter:
\begin{equation} \label{varianza}
Var(a) = \mathbb E(W_{a,b})^2 - (\mathbb E(W_{a,b})) ^2 \approx a^\beta
\end{equation}
where $\beta \in [-1,3]$.
\cite{malamud1999}  find a relationship between $H$ and $\beta$ for self-affine series. Consequently, the Hurst exponent is defined as: 

\begin{itemize}
	\item $H=\frac{\beta+1}{2}$, with $\beta \in [-1,1) $, if the signal is a fGn,   
	\item $H=\frac{\beta-1}{2}$,  with $\beta \in [1,3]$, if the signal is a  fBm,
\end{itemize}
where fBm stands for fractional Brownian motion, and fGn for fractional Gaussian noise.

The wavelet method used in this paper is a modification of the AWC method developed in \cite{shn98}, who propose the following steps:
\begin{enumerate}
	\item Apply the wavelet transform of the data in the wavelet domain, $W_{a,b}$;
	\item Compute, for a fixed scale, $a$, the average wavelet coefficients; 
	\item Draw the log-log plot of coefficients vs. scale $a$. 
\end{enumerate}
  
The key improvement presented in \cite{pa2017} consists in the use of more robust estimators for the computation of the coefficients required in step 2, compared to the estimators used in \cite{shn98}.

Based on Eq. (\ref{eq-tw}), it is proposed to estimate the coefficients of the variance of the time series using two estimators. The first one is the well-known unbiased variance estimator $\widehat{Var}$, which given a set of data $x_1,\ldots,x_m$ is defined as:
\begin{equation}
\widehat{Var}(x)=\frac{1}{m-1}\sum_{i=1}^{m} (x_i - \bar{x})^2
\end{equation} 
The second one is the median absolute deviation (MAD), which is a more robust estimator of the variance (see \cite{maronna2006}):
\begin{equation}\label{mad}
	MAD(W_{a,b})=Med(|W_{a,b}-Med(W_{a,b})|)
\end{equation} 
where $Med(\cdot)$ is the median operator. This improved method will be referred, hereinafter, as AWC-MAD. In \cite{pa2017} AWC-MAD was benchmarked against the classical R/S method and competing alternatives for averaging wavelet coefficients. In all cases AWC-MAD estimations of the Hurst exponent were closer to the theoretical Hurst exponent in synthetic series with 32768 datapoints. Subsequently, \cite{pa2018,pa2019,ICIAM2019} compare R/S and AWC-MAD methods for climate time series as well as for synthetic time series.

Considering that the main goal of this paper is to study long range dependence in the cryptocurrency market, with shorter time series, we benchmark theoretical and estimated Hurst exponent on fBm and fGn sinthetic series with 1435 datapoints. We generate rolling windows of 500 observations, moving forward one observation in each window. The algorithm to generate the artificial series was proposed by \cite{abry1996}, and implemented in MatLab with the function \texttt{wfbm.m}, using Daubechies wavelet of order 10. Results displayed in Table \ref{tab:sinthetic-results}. The first column is the theoretical Hurst exponent value, and the other columns display the mean and standard deviation of the estimations computed on all the rolling windows with the AWC-MAD.

\begin{table}[htbp]
  \centering
\caption{Estimation of the Hurst exponent for fBm and fGn sinthetic series, using the AWC-MAD method.}
    \begin{tabular}{ccccc}
    \toprule
          & \multicolumn{2}{c}{Estimated H (fBm series)} & \multicolumn{2}{c}{Estimated H (fGn series)} \\
    \multicolumn{1}{l}{Theoretical H} & \multicolumn{1}{c}{Mean} & \multicolumn{1}{c}{Std. Dev.} & \multicolumn{1}{c}{Mean} & \multicolumn{1}{c}{Std. Dev.} \\
    \midrule
    0.2   & 0.2649 & 0.0712 & 0.2602 & 0.0117 \\
    0.4   & 0.3513 & 0.0425 & 0.4097 & 0.0274 \\
    0.6   & 0.5697 & 0.0501 & 0.6044 & 0.0280 \\
    0.8   & 0.7552 & 0.0683 & 0.7898 & 0.0321 \\
    \bottomrule
    \end{tabular}%
\label{tab:sinthetic-results}
\end{table}%

\section{Data and empirical results\label{sec:application}}
Cryptocurrencies are traded in online platforms that are not compelled to abide by national financial regulations. Thus, a careful selection of a reliable data source is crucial for the validity of results. Following \cite{Alexander2020} and \cite{BARIVIERA2020}, we downloaded data from \cite{cryptocompare}. We use high frequency (every two hour) price data of eleven important cryptocurrencies in terms of to traded volume. The period under examination goes from 14/11/2019 to 08/06/2020, for a total of 2496 observations. 
We studied two-hour returns and volatility. Considering that there are several ways of capturing volatility in financial markets, we selected two widely used proxies \cite{Gallant1999,Alizadeh2002,Cotter2011}. Additionally, in order to take into account the dynamic character of the market, we compute the Hurst exponents by means of rolling windows. Each rolling window cointains 360 datapoints (30 trading days). Each window moves one observation forward, and deletes the first observation. Data covers the period that lies before and after the onset of the Covid-19 pandemic.
Namely, we compute the following measures from the price time series: 
\begin{itemize}
\item Logarithmic or continuously compounded return:
\begin{equation}R_t=\log(P_t)-\log(P_{t-1}) \end{equation}
where $P_t$ and $P_{t-1}$ are two consecutive closing prices every two hours. 
\item  Max-Min volatility:
\begin{equation}Vol_t^{MinMax}=\log(P_t^{max})-\log(P_t^{min})\end{equation}
where $P_t^{max}$ and $P_t^{min}$ the maximum and minimum prices observed in a two-hour period.
\item Absolute return volatility:
\begin{equation}Vol_t^{Abs}=abs(\log(P_t)-\log(P_{t-1})) \end{equation}
\end{itemize}

\begin{figure}[!htbp]
\centering
\subfloat{\includegraphics[width = 0.33\textwidth]{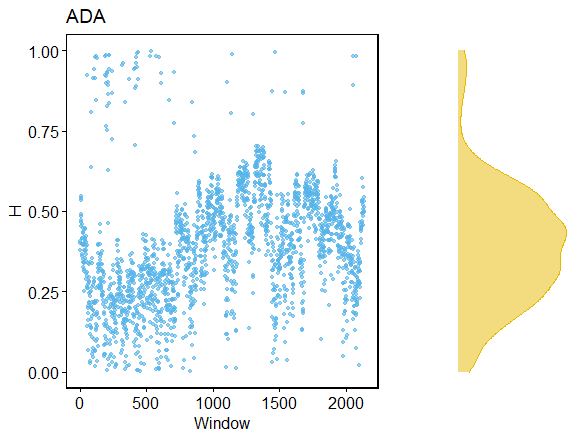}}
\subfloat{\includegraphics[width = 0.33\textwidth]{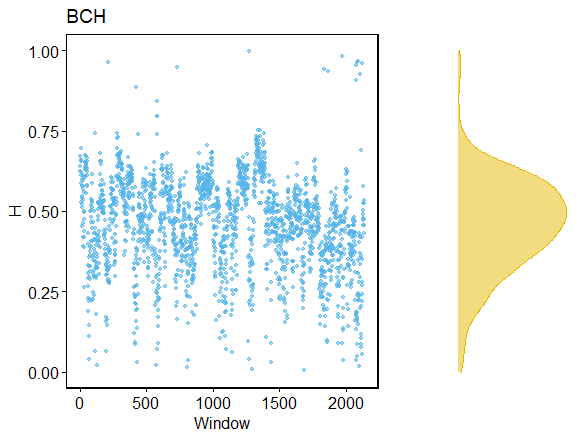}}
\subfloat{\includegraphics[width = 0.33\textwidth]{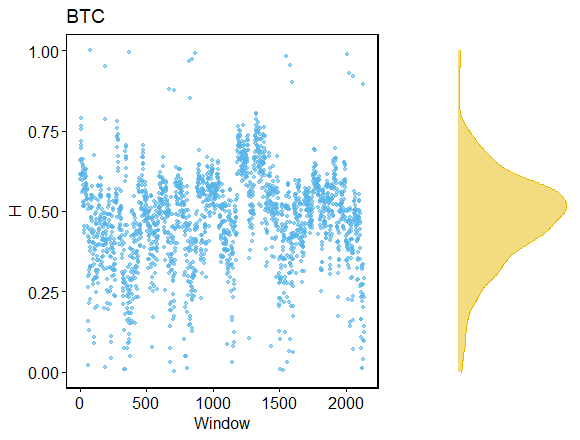}}\\ 
\subfloat{\includegraphics[width = 0.33\textwidth]{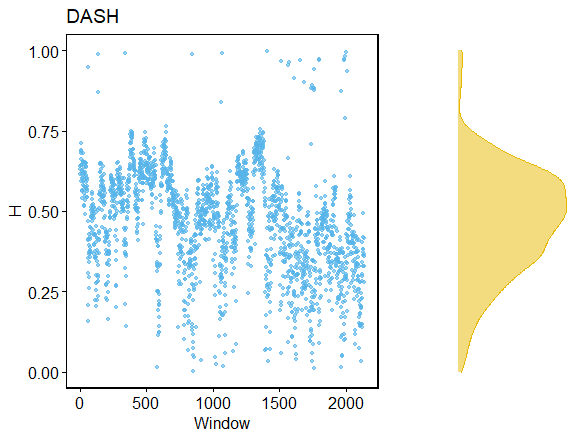}}
\subfloat{\includegraphics[width = 0.33\textwidth]{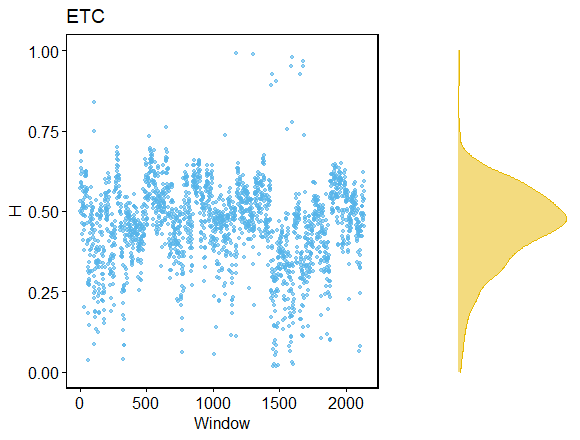}}
\subfloat{\includegraphics[width = 0.33\textwidth]{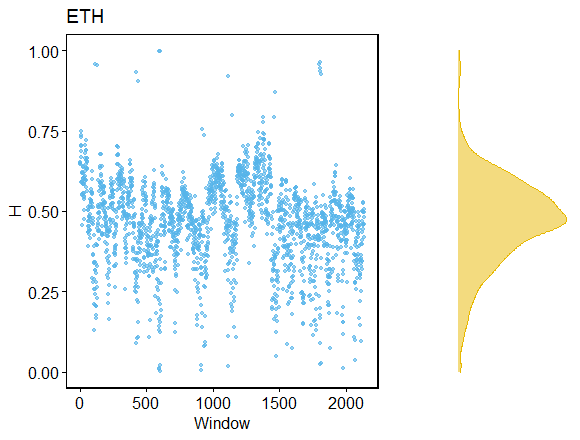}}\\
\subfloat{\includegraphics[width = 0.33\textwidth]{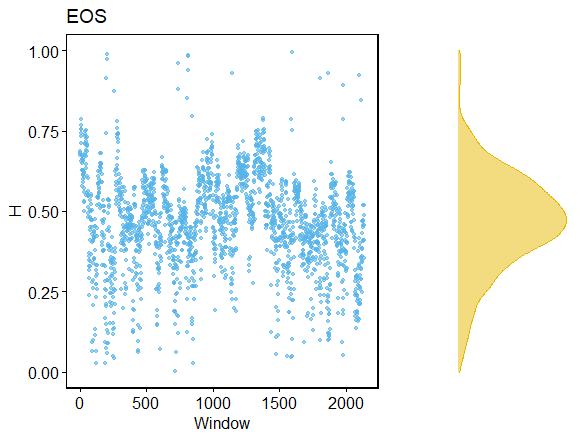}}
\subfloat{\includegraphics[width = 0.33\textwidth]{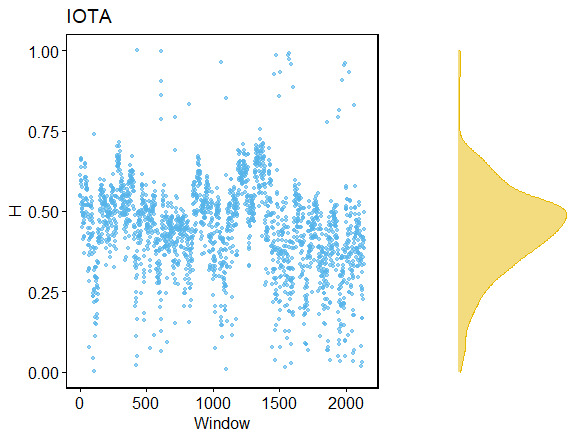}}
\subfloat{\includegraphics[width = 0.33\textwidth]{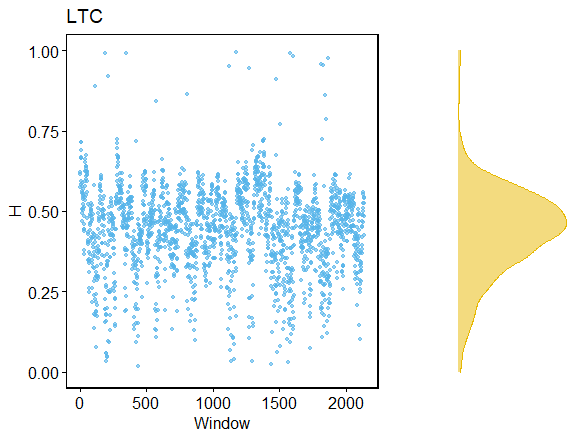}}\\
\subfloat{\includegraphics[width = 0.33\textwidth]{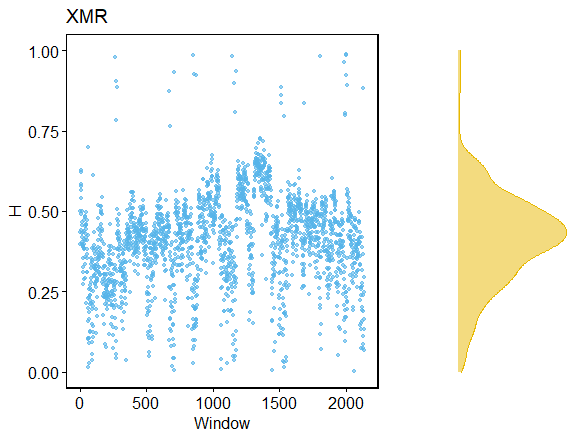}}
\subfloat{\includegraphics[width = 0.33\textwidth]{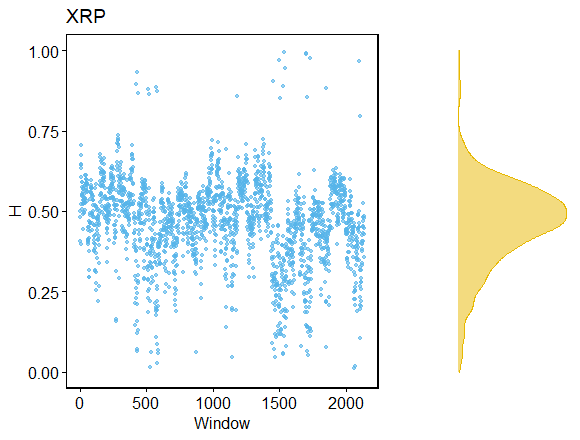}}
\caption{Hurst exponent of two-hour returns, using rolling windows.}
\label{fig:HurstReturns}
\end{figure}

We obtained important results regarding long memory in returns and in volatility. With respect to returns, as can be appreciated in Figure \ref{fig:HurstReturns}, the effect of Covid-19 on the Hurst exponent is mild. The average Hurst exponent around 0.5, which is similar to previous findings for time series at low and high frequency sampling \citep{Bariviera2017,Bariviera2018}. Table \ref{tab:HurstReturn} displays the mean Hurst exponent of each return time series, for all windows, windows before 03/03/2020, windows during two weeks in March when there was a peak in the pandemic, and windows after 18/03/2020. We observe that between March 3rd. and March 18th. the Hurst exponents are greater than in the preceding period, but they recover previous levels soon after this date.  

\begin{figure}[!htbp]
\centering
\subfloat{\includegraphics[width = 0.33\textwidth]{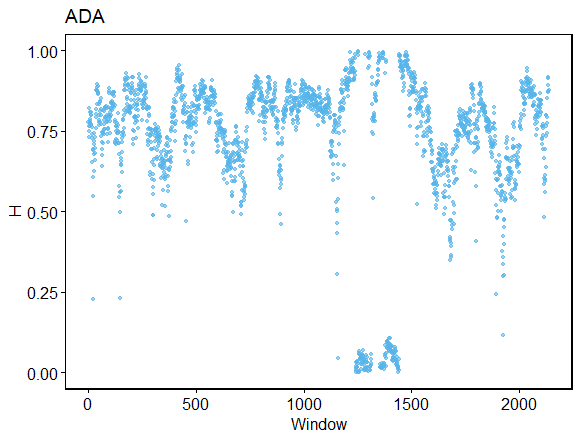}} 
\subfloat{\includegraphics[width = 0.33\textwidth]{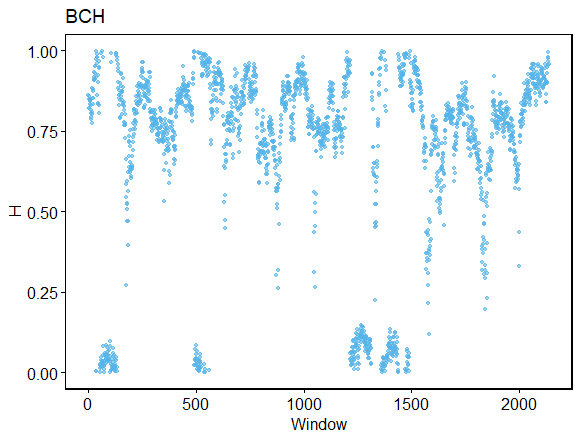}}
\subfloat{\includegraphics[width = 0.33\textwidth]{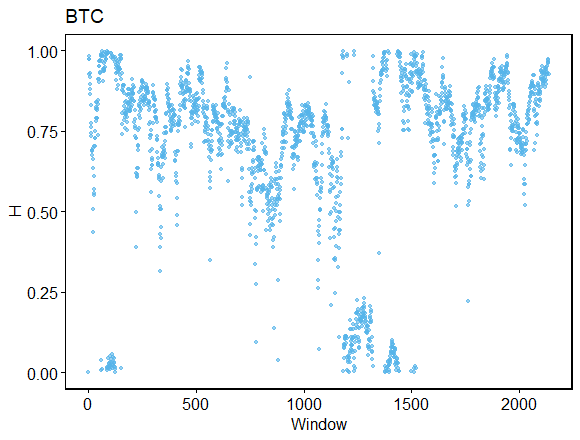}}\\
\subfloat{\includegraphics[width = 0.33\textwidth]{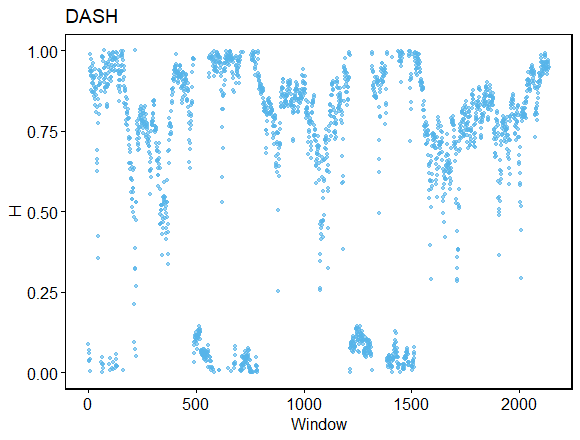}}
\subfloat{\includegraphics[width = 0.33\textwidth]{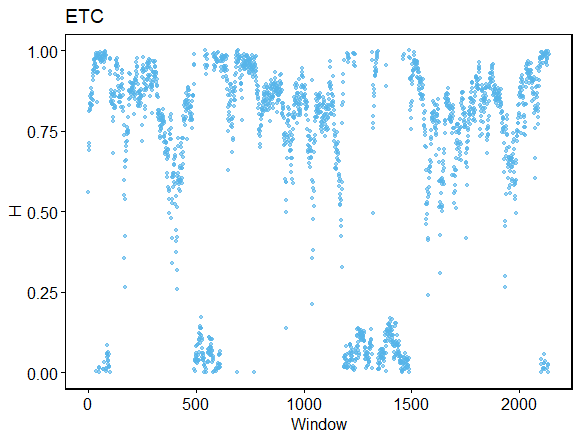}}
\subfloat{\includegraphics[width = 0.33\textwidth]{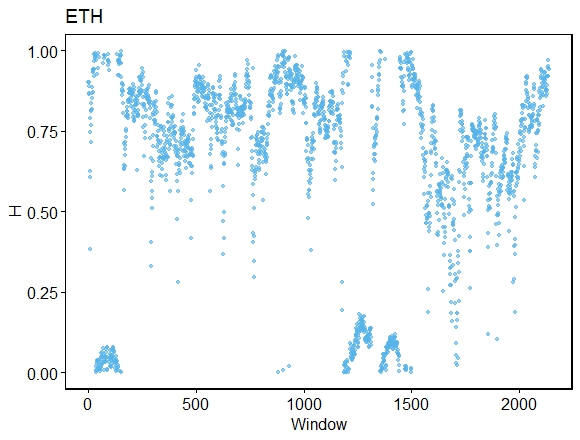}}\\
\subfloat{\includegraphics[width = 0.33\textwidth]{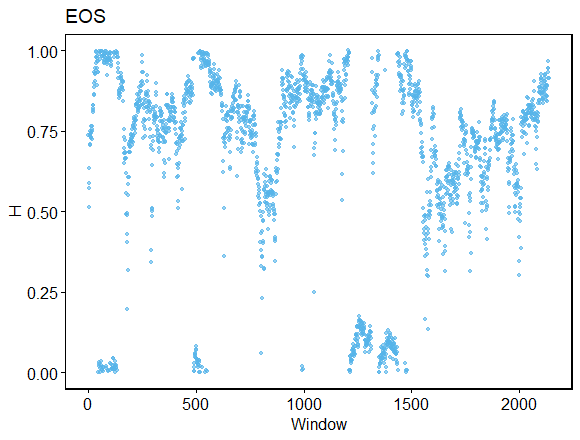}}
\subfloat{\includegraphics[width = 0.33\textwidth]{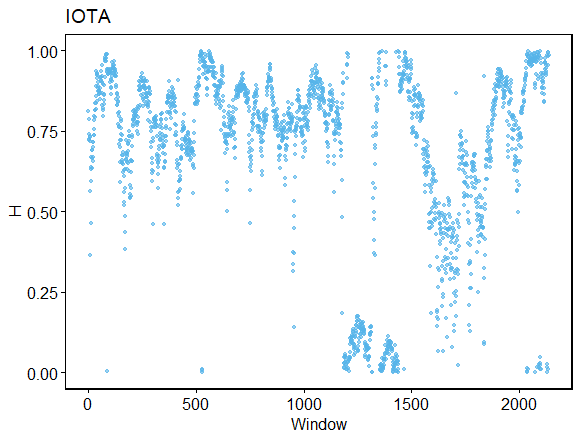}}
\subfloat{\includegraphics[width = 0.33\textwidth]{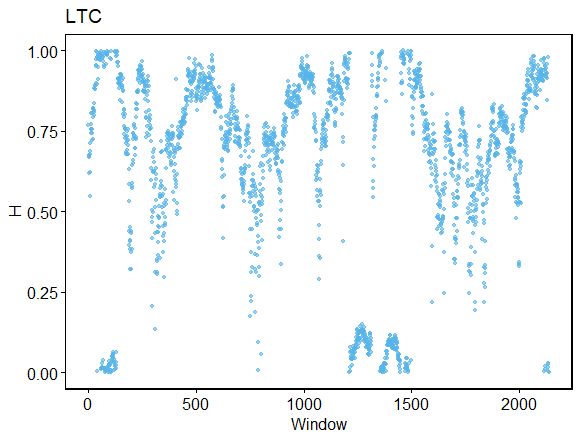}}\\
\subfloat{\includegraphics[width = 0.33\textwidth]{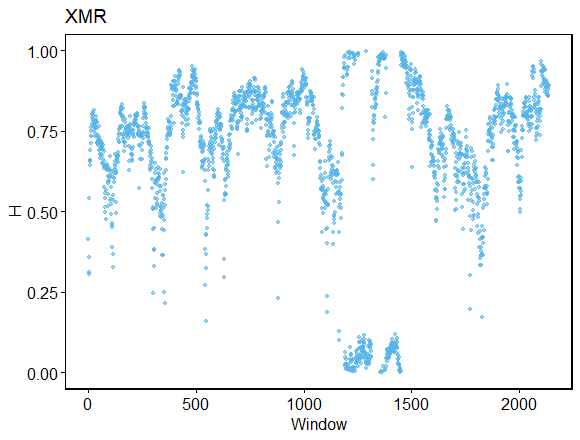}}
\subfloat{\includegraphics[width = 0.33\textwidth]{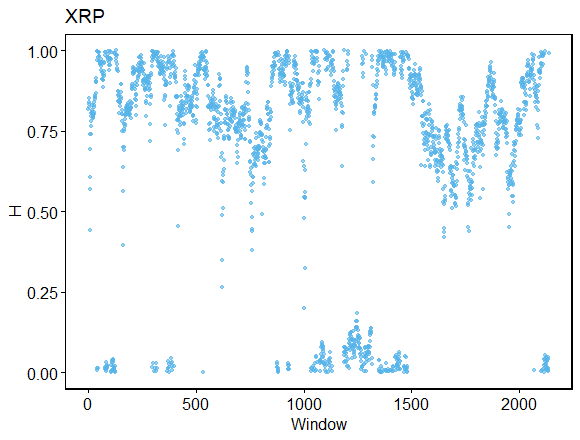}}
\caption{Hurst exponent of Max-Min two-hour volatility, using rolling windows.}
\label{fig:HurstMaxMin}
\end{figure}

A different picture is obtained when we analyze volatility. There has been a striking effect of Covid-19 on the long-term memory of volatility. This paper uses two proxies for volatility. The first one, is the Max-Min volatility, and the dynamic evolution of its Hurst exponent is displayed in Figure \ref{fig:HurstMaxMin}. Considering that this way of measuring volatility uses the highest and lowest price of a given cryptocurrency within a two-hour period, it is suitable to capture extreme events.  
In Table \ref{tab:HurstMaxMin} we observe that the period that goes from November to the early March exhibits very highly persistence (Hurst exponent greater than 0.69). On the contrary, between March 3rd. and March 18th. the Hurst exponents drop suddendly, reflecting an antipersistent time series. This situation could reflect a moment of market panic, where large positive returns are followed by large negative returns and vice versa. We should recall that during these days stocks markets around the world were in free fall (for example, the German DAX index plunged by 26\%). Cryptocurrencies, as alternative assets, were subject to speculative moves, and those large swings in volatilities could reflect the uncertainty around the true value and feasibility of cryptocurrencies as a safe haven amid the pandemic. After March 18th. the Hurst values reversed to values above 0.7, returning volatility into a highly persisting process.

\begin{figure}[!htbp]
\centering
\subfloat{\includegraphics[width = 0.33\textwidth]{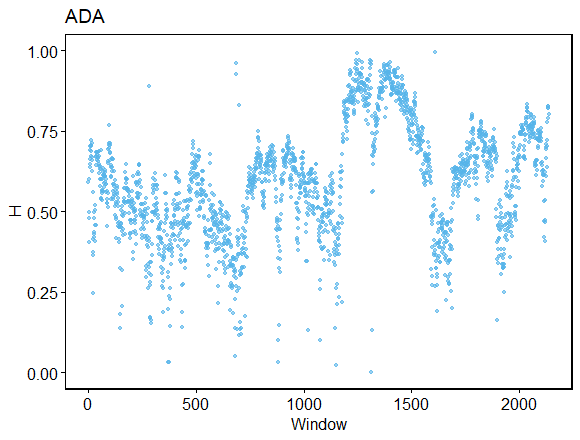}} 
\subfloat{\includegraphics[width = 0.33\textwidth]{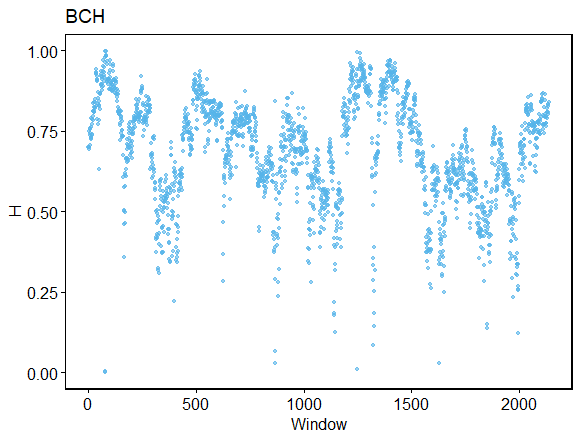}}
\subfloat{\includegraphics[width = 0.33\textwidth]{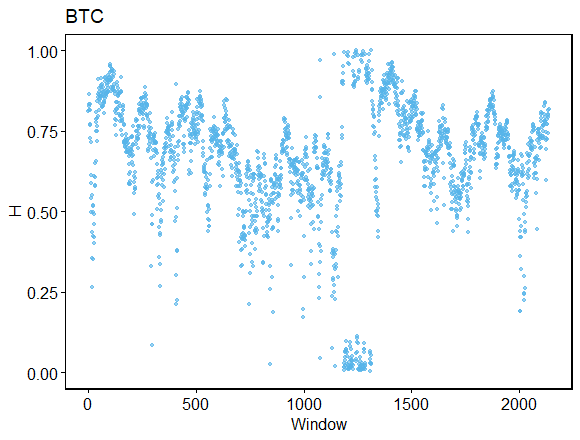}}\\
\subfloat{\includegraphics[width = 0.33\textwidth]{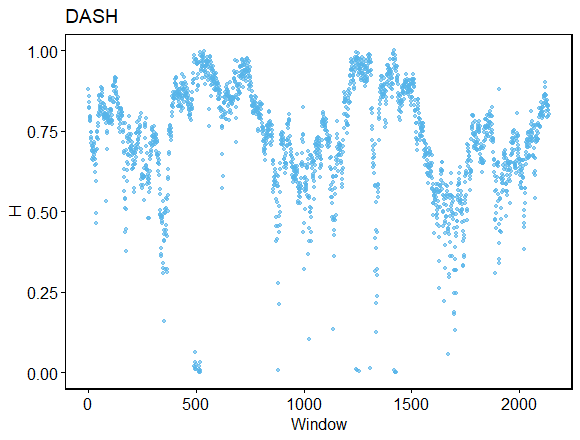}}
\subfloat{\includegraphics[width = 0.33\textwidth]{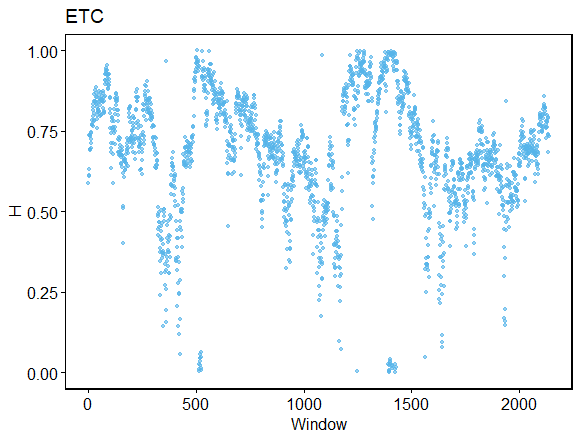}}
\subfloat{\includegraphics[width = 0.33\textwidth]{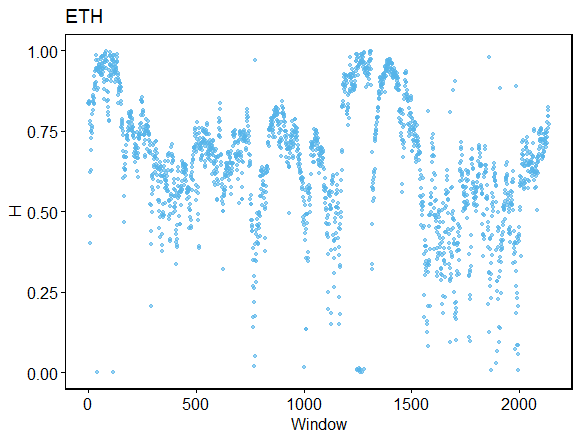}}\\
\subfloat{\includegraphics[width = 0.33\textwidth]{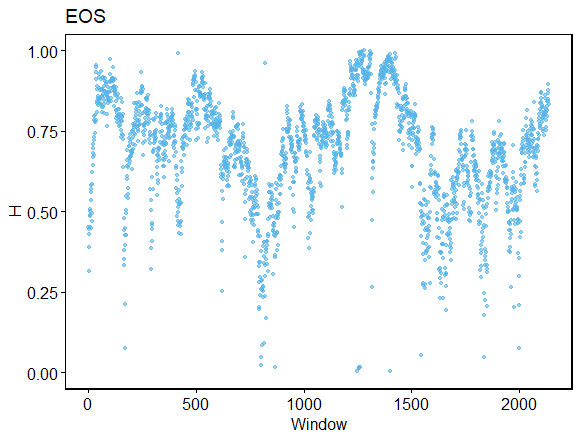}}
\subfloat{\includegraphics[width = 0.33\textwidth]{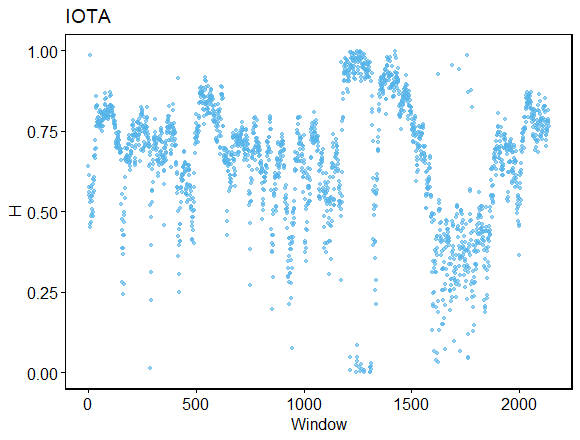}}
\subfloat{\includegraphics[width = 0.33\textwidth]{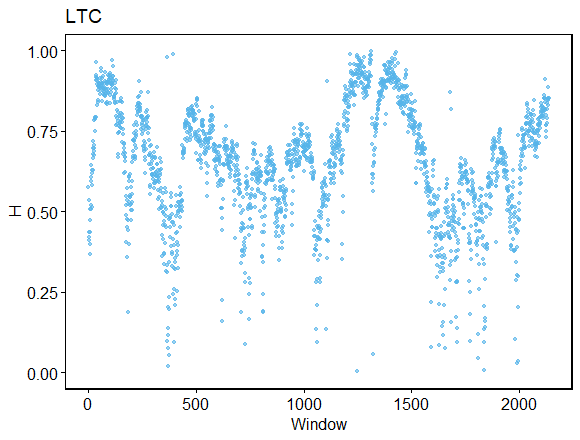}}\\
\subfloat{\includegraphics[width = 0.33\textwidth]{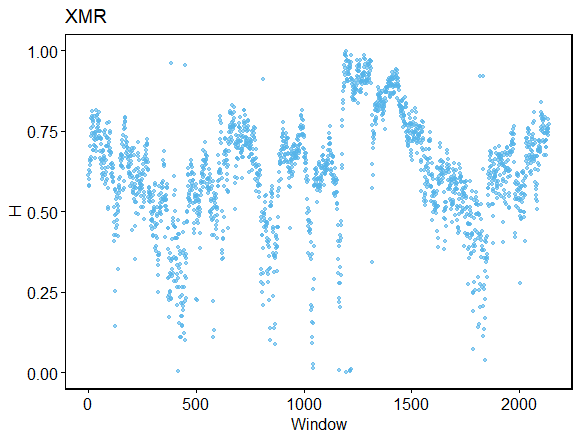}}
\subfloat{\includegraphics[width = 0.33\textwidth]{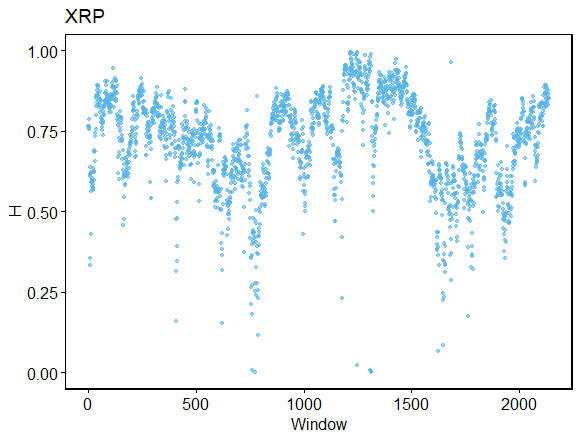}}
\caption{Hurst exponent of two-hour absolute returns, using rolling windows.}
\label{fig:HurstAbsoluteReturns}
\end{figure}

Absolute returns have been proposed to measure financial time series volatility. Similar to Max-Min volatility, absolute returns also exhibit a persistent behavior during the period before March. However, and at odds with the previous proxy, absolute returns increased in their persistence during the upheaval of the Covid-19 crisis (See Figure \ref{fig:HurstAbsoluteReturns} and Table \ref{tab:HurstAbsoluteReturns}). 

These two apparently contradictory findings could be reconciled as follows. First, there is undoubtedly a strong Covid-19 effect on volatility, as both proxies reflect strong changes in their Hurst exponents. Second, Max-Min and absolute returns measure different aspects of volatility. The former is more sensible to extreme events, considers only the positive side of volatility, and measures the range of prices during a given hour. The latter, conversely, considers both positive and negative changes as single absolute returns, being less sensitive to extreme events. Analyzing both results jointly, we can conclude that hours with large price swings were followed by hours with smaller price swings during the worst days of the pandemic. However, two-hour returns were more persistent during the same period. After March 18th. extreme events (as measured by Max-Min volatility) became again to be highly persistent, along with a persistence (albeit to a lesser extent) of absolute returns. 

\begin{table}[htbp]
  \centering
  \caption{Mean Hurst exponent of return time series, for all windows, windows before 03/03/2020, windows between 03/03/2020 and 18/03/2020, and windows after 18/03/2020}
 \resizebox{.99\textwidth}{!}{
   \begin{tabular}{lrrrr}
\cmidrule{1-5}          & \multicolumn{4}{c}{Hurst exponent} \\
          & \multicolumn{1}{l}{All windows} & \multicolumn{1}{l}{Before 03/03/2020} & \multicolumn{1}{l}{Between 03 and 18/03/2020} & \multicolumn{1}{l}{After 18/03/2020} \\
    \midrule
    ADA   & 0.3869 & 0.3538 & 0.4727 & 0.4259 \\
    BCH   & 0.4589 & 0.4727 & 0.5198 & 0.4092 \\
    BTC   & 0.4747 & 0.4657 & 0.5529 & 0.4664 \\
    DASH  & 0.4662 & 0.5014 & 0.5321 & 0.3706 \\
    EOS   & 0.4655 & 0.4755 & 0.5430 & 0.4179 \\
    ETC   & 0.4576 & 0.4711 & 0.4409 & 0.4353 \\
    ETH   & 0.4670 & 0.4790 & 0.5203 & 0.4234 \\
    IOTA  & 0.4448 & 0.4666 & 0.5089 & 0.3772 \\
    LTC   & 0.4433 & 0.4461 & 0.4778 & 0.4257 \\
    XMR   & 0.4085 & 0.3914 & 0.5061 & 0.4102 \\
    XRP   & 0.4656 & 0.4855 & 0.4632 & 0.4253 \\
    \bottomrule
    \end{tabular}%
		}
  \label{tab:HurstReturn}%
\end{table}%

\begin{table}[htbp]
  \centering
  \caption{Mean Hurst exponent of Max-Min volatility time series, for all windows, windows before 03/03/2020, windows between 03/03/2020 and 18/03/2020, and windows after 18/03/2020}
 \resizebox{.99\textwidth}{!}{
   \begin{tabular}{lrrrr}
\cmidrule{1-5}          & \multicolumn{4}{c}{Hurst exponent} \\
          & \multicolumn{1}{l}{All windows} & \multicolumn{1}{l}{Before 03/03/2020} & \multicolumn{1}{l}{Between 03 and 18/03/2020} & \multicolumn{1}{l}{After 18/03/2020} \\
\cmidrule{1-5}    ADA   & 0.7278 & 0.7593 & 0.5529 & 0.7231 \\
    BCH   & 0.6907 & 0.7005 & 0.4626 & 0.7493 \\
    BTC   & 0.7050 & 0.6738 & 0.5894 & 0.8097 \\
    DASH  & 0.6732 & 0.6625 & 0.4485 & 0.7731 \\
    EOS   & 0.6814 & 0.7138 & 0.4833 & 0.6825 \\
    ETC   & 0.6794 & 0.6994 & 0.2783 & 0.7766 \\
    ETH   & 0.6597 & 0.6954 & 0.4468 & 0.6593 \\
    IOTA  & 0.6895 & 0.7399 & 0.4887 & 0.6543 \\
    LTC   & 0.6645 & 0.6911 & 0.4162 & 0.6952 \\
    XMR   & 0.6883 & 0.6927 & 0.5452 & 0.7284 \\
    XRP   & 0.7114 & 0.7214 & 0.5960 & 0.7306 \\
    \bottomrule
    \end{tabular}%
		}
  \label{tab:HurstMaxMin}%
\end{table}%

\begin{table}[htbp]
  \centering
  \caption{Mean Hurst exponent of absolute return time series, for all windows, windows before 03/03/2020, windows between 03/03/2020 and 18/03/2020, and windows after 18/03/2020}
 \resizebox{.99\textwidth}{!}{
  \begin{tabular}{lrrrr}
\cmidrule{1-5}          & \multicolumn{4}{c}{Hurst exponent} \\
          & \multicolumn{1}{l}{All windows} & \multicolumn{1}{l}{Before 03/03/2020} & \multicolumn{1}{l}{Between 03 and 18/03/2020} & \multicolumn{1}{l}{After 18/03/2020} \\
    \midrule
    ADA   & 0.5920 & 0.5420 & 0.8421 & 0.6093 \\
    BCH   & 0.6874 & 0.7078 & 0.8087 & 0.6033 \\
    BTC   & 0.6651 & 0.6367 & 0.7892 & 0.6811 \\
    DASH  & 0.7258 & 0.7493 & 0.8206 & 0.6443 \\
    EOS   & 0.6882 & 0.7084 & 0.8499 & 0.5906 \\
    ETC   & 0.6825 & 0.6937 & 0.8012 & 0.6184 \\
    ETH   & 0.6518 & 0.6825 & 0.8446 & 0.5216 \\
    IOTA  & 0.6578 & 0.6824 & 0.8012 & 0.5572 \\
    LTC   & 0.6536 & 0.6588 & 0.8549 & 0.5730 \\
    XMR   & 0.6233 & 0.6049 & 0.8319 & 0.5896 \\
    XRP   & 0.7182 & 0.7271 & 0.8460 & 0.6555 \\
    \bottomrule
    \end{tabular}%
		}
  \label{tab:HurstAbsoluteReturns}%
\end{table}%

\section{Conclusions \label{sec:conclusions}}

This paper's contribution is twofold. On the one hand, it applies a new, improved, wavelet-based method to compute the Hurst exponent. On the other hand, it provides an analysis of high frequency return and volatility of eleven cryptocurrencies during Covid-19 pandemic. We find that, even though the ongoing pandemic has produced only a mild effect on the long-range memory of cryptocurrency returns, it imprinted a strong transitory effect on volatility. We use two alternative measures of return volatility: Max-Min and absolute returns. Both proxies reflect a significant change in their long memory profile. Regarding Max-Min, we observe a momentary swap, from a highly persistent process to an antipersistent one. Differently, absolute returns time series suffer a reinforcement in their persistent behavior during the upheaval of the pandemic. The divergent reaction of both volatility measures could be explained by the sensibility of the Max-Min measure to more extreme events, as it only measures positive changes ($Vol_t^{MinMax}=\log(P_t^{max})-\log(P_t^{min})$). The Hurst exponent of both volatility proxies return to levels previous to the Covid-19 pandemic, after March 18th. Thus, we conclude that this one-time-only event has had only a transitory effect on the long memory profile of returns and volatilities.


\bibliographystyle{apalike}
\bibliography{biblio}

\end{document}